\documentclass[prb,aps,twocolumn,showpacs,preprintnumbers,amsmath,amssymb]{revtex4}

\usepackage{graphicx}
\usepackage{dcolumn}

\begin{document}

\title{Band symmetries of mixed-valence topological insulator: SmB$_{6}$}

\author{Chang-Jong Kang$^1$, Junwon Kim$^1$, Kyoo Kim$^1$,
J.-S. Kang$^2$, J. D. Denlinger$^3$, and B. I. Min$^1$}
\affiliation{
$^1$Department of Physics, PCTP,
	Pohang University of Science and Technology,
	Pohang 790-784, Korea\\
$^2$Department of Physics, The Catholic University of Korea,
	Bucheon 420-743, Korea\\
$^3$Advanced Light Source, Lawrence Berkeley Laboratory,
	Berkeley, CA 94720, U.S.A.
}
\date{\today}

\begin{abstract}
We have investigated the band structure and the band symmetry of
mixed-valence insulator SmB$_{6}$
systematically within the density functional theory (DFT).
We have analyzed the symmetries and characters of
Sm $4f$ and $5d$ bands near the Fermi level ($E_F$)
in terms of the non-relativistic real cubic bases
as well as the relativistic complex bases
incorporating the spin-orbit coupling and the cubic crystal field.
Further, we have found that the semi-core band located
at $\sim 15$ eV below $E_F$ has the mixed parity and thereby
affects the parity eigenvalues of the special ${\bf k}$-points considerably.
We have discussed the possible surface states and topological
class of SmB$_{6}$, based on the bulk parity tables.
\end{abstract}

\pacs{71.15.Mb, 71.20.-b, 71.27.+a, 73.20.At}

\maketitle

\section{Introduction}

SmB$_{6}$ has been studied for decades as a typical
mixed-valence system~\cite{Cohen70,Campagna79,Eibschutz72,Chazalviel76,Beaurepaire90}
and also as a Kondo insulator.\cite{Mandrus94,Cooley95,Demsar95}
Research on SmB$_{6}$ has been revived owing to the
recent suggestion of it to be a topological Kondo insulator.
\cite{Dzero10,Takimoto11,Dzero12,Feng13,Alexandrov13}
Then various follow-up experiments were actively carried out
to verify this issue.\cite{Xu13,Neupane13,Kim13,Kim13-2}
The topological insulator can be identified
by the existence of the in-gap states with a Dirac-cone dispersion,
and so several angle-resolved
photoemission spectroscopy (ARPES) experiments were reported
to probe the in-gap states.\cite{Miyazaki12,Xu13,Neupane13,JD13}
However, SmB$_{6}$ has a very narrow band gap and a large effective mass
due to the hybridization between the localized $f$-electron
and conduction electrons, which makes it difficult
to observe the clear in-gap states near the Fermi level ($\rm E_F$)
within the limit of the current experimental resolution.

Topological insulators are classified by four $\mathbb{Z}_{2}$
topological indices $\nu_{0};(\nu_{1}\nu_{2}\nu_{3})$,
which distinguish between strong ($\nu_{0} = 1$) and weak ($\nu_{0} = 0$)
topological insulators and give information
on the surface topology protected by the time reversal
symmetry.\cite{Fu07prl,Hasan10}
The number of Dirac points is odd on any surface
of a strong topological insulator,
while it is even or zero for a weak topological insulator.
If the system has the inversion symmetry,
four $\mathbb{Z}_{2}$ topological indices can be easily obtained
from the products of parity eigenvalues.\cite{Fu07}
Hence, the information on the surface topology can be inferred
from the bulk band parity eigenvalues at the high symmetry ${\bf k}$-points.
Therefore, in the theoretical aspect,
it is important to know the detailed characters and symmetries
of bulk band structures.

For SmB$_{6}$, there have been a few {\it ab initio} band structure calculations.
But the systematic analysis of band symmetry, which is important in examining the
topological symmetry, is lacking.
In this article, we have investigated electronic structure of SmB$_{6}$
and determined the band symmetries of its bulk band structure
within the density functional theory (DFT).
Interestingly, we have found a semi-core band having the mixed parity,
which affects the parity eigenvalues of the special ${\bf k}$-points
in the bulk Brillouin zone considerably.
We have discussed the topological class of SmB$_{6}$,
and provided possible surface states based on the bulk parity table.

\begin{figure}[b]
\includegraphics[width=8 cm]{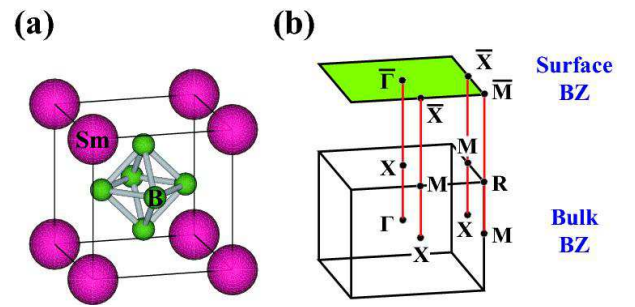}
\caption{(Color Online)
(a) Crystal structure of SmB$_{6}$.
(b) The high symmetry ${\bf k}$-points in the bulk Brillouin zone,
and the time-reversal invariant momentum (TRIM) points
in the (001) surface Brillouin zone.
}
\label{structure}
\end{figure}

\section{Computations}

We have employed the full-potential linearized augmented plane wave (FLAPW) band method
implemented in Wien2k~\cite{Wien2k} for the analysis of band structures.
For the structural relaxation, we have also used the projector augmented wave (PAW)
band method implemented in VASP.\cite{VASP}
For the exchange-correlation, we have utilized the generalized gradient approximation
(GGA) of Perdew, Burke and Ernzerhof.
The spin-orbit coupling (SOC) is taken into account in the second variational scheme.
For the band structures, we have used a 17 $\times$ 17 $\times$ 17 k-point mesh
in the full Brillouin zone.
The muffin-tin radii in the FLAPW method were set to 2.50 a.u., 1.54 a.u.
for Sm, and B, respectively and the product of the muffin-tin radius
and the maximum reciprocal lattice vector $K_{max}$, $R_{MT}\cdot K_{max} = 7$.
The maximum $L$ value for the waves inside the atomic spheres, $L_{max} = 10$,
and the largest $G$ in the charge Fourier expansion $G_{max} = 12$
were used in the calculations.

SmB$_{6}$ has the simple cubic structure (space group: Pm$\bar{3}$m),
as shown in Fig.~\ref{structure}(a).
First, we have fully relaxed structural parameters of SmB$_{6}$ in the GGA + SOC using the VASP.
We have stopped the relaxation when forces exerted on every atom are smaller than 1 meV/$\AA$.
Resulting structure has the lattice constant $a$ = 4.1062 $\AA$
and the atomic position parameter of boron $x$ = 0.1992,
which are quite comparable to the experimental data $a$ = 4.1327 $\AA$
and $x$ $\sim$ 0.2 at 100 K.\cite{Funahashi10}

\section{Results}
\subsection{Band structure and symmetry}

\begin{figure}[b]
\includegraphics[width=8 cm]{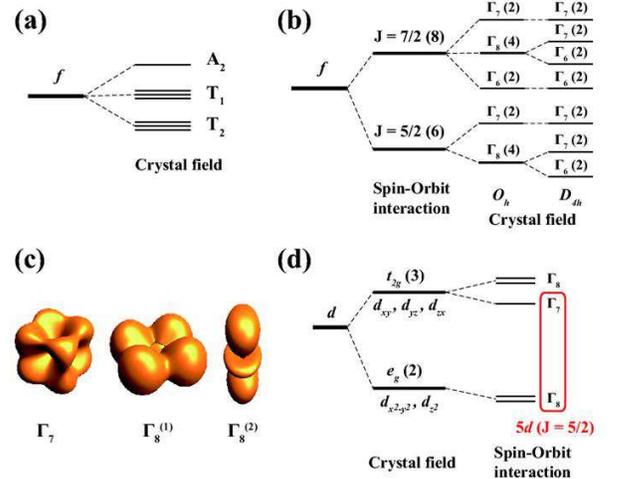}
\caption{(Color Online)
Crystal field splittings.
(a) Sm 4$f$ levels in the absence of SOC are split into $A_{2}$, $T_{1}$, and $T_{2}$
states in the cubic symmetry.
(b) Sm 4$f$ levels in the presence of SOC are split first into
$J = 5/2$ and $J = 7/2$ states, which are split further
under the crystal fields of cubic ($O_{h}$) and tetragonal ($D_{4h}$) symmetry.
The bases are represented by Bethe's relativistic double group notations.
(c) Shapes of wave functions of $\Gamma_{7}$ and $\Gamma_{8}$
belonging to $J = 5/2$ states.
Under the crystal fields of tetragonal ($D_{4h}$) symmetry,
$\Gamma_{8}^{(1)}$ and $\Gamma_{8}^{(2)}$ in cubic symmetry
turn into $\Gamma_{7}$ and $\Gamma_{6}$, respectively.
(d) Crystal field and the SOC splittings for Sm 5$d$ levels.
$\Gamma_{8}$ of $e_{g}$ states
will be split into $\Gamma_{7}$ ($d_{x^2-y^2}$) and
$\Gamma_{6}$ ($d_{z^2}$), respectively, in tetragonal symmetry.
}
\label{cef}
\end{figure}

Sm ion in SmB$_{6}$ is surrounded by B cages located in each corner of the
cubic cell, and so
electrons in the Sm ion feel the cubic crystal field.
As a result, in the absence of the SOC, Sm $f$ states
are split into three real cubic harmonic bases,
$A_{2}$, $T_{1}$, and $T_{2}$ states, as shown in Fig.~\ref{cef}(a),
which are given by:
\begin{equation}
\begin{split}
&T_{1}(x) = \sqrt{\frac{7}{16\pi}} x(5x^2-3r^2)/r^3,
\\
&T_{1}(y) = \sqrt{\frac{7}{16\pi}} y(5y^2-3r^2)/r^3,
\\
&T_{1}(z) = \sqrt{\frac{7}{16\pi}} z(5z^2-3r^2)/r^3,
\\
&T_{2}(\xi) = \sqrt{\frac{105}{16\pi}} x(z^2-y^2)/r^3,
\\
&T_{2}(\eta) = \sqrt{\frac{105}{16\pi}} y(z^2-x^2)/r^3,
\\
&T_{2}(\zeta) = \sqrt{\frac{105}{16\pi}} z(x^2-y^2)/r^3,
\\
&A_{2} = \sqrt{\frac{105}{16\pi}} xyz/r^3.
\end{split}
\label{harmonics}
\end{equation}

However, Sm $f$-electrons feel much larger SOC than the cubic crystal field.
Then, instead of the real cubic harmonic bases,
one needs to use the relativistic double group bases that incorporate
both the SOC and the cubic crystal field~\cite{Pappalardo61}
(see the Bethe notations for the relativistic crystal field splitting
in Fig.~\ref{cef}(b)).
For Sm$^{3+}$ ion, $f$ states are split into
lower $J = 5/2$ and upper $J = 7/2$ states,
and the lower $J = 5/2$ states in the presence of cubic crystal field are split
further into $\Gamma_{7}^{-}$ doublet and $\Gamma_{8}^{-}$ quartet
(here $-$ in the superscript denotes the negative parity),
which are given by
\begin{equation}
\begin{split}
&\Gamma_{7} = \sqrt{\frac{5}{6}}\bigg| J_{z}=\pm\frac{3}{2}\bigg\rangle
- \sqrt{\frac{1}{6}}\bigg| J_{z}=\mp\frac{5}{2}\bigg\rangle,
\\
&\Gamma_{8}^{(1)} = \sqrt{\frac{1}{6}}\bigg| J_{z}=\pm\frac{3}{2}\bigg\rangle
+ \sqrt{\frac{5}{6}}\bigg| J_{z}=\mp\frac{5}{2}\bigg\rangle,
\\
&\Gamma_{8}^{(2)} = \bigg| J_{z}=\pm\frac{1}{2}\bigg\rangle.
\end{split}
\label{j=5/2}
\end{equation}
As shown in Fig.~\ref{cef}(c),
$\Gamma_{7}$ doublet have lobes along the corners of the cubic lattice,
whereas $\Gamma_{8}$ quartet have lobes along the axial directions:
$\Gamma_{8}^{(1)}$ along \emph{x} and \emph{y}-axes and
$\Gamma_{8}^{(2)}$ along the \emph{z}-axis.
On the other hand,
the upper $J = 7/2$ states in the cubic crystal field are split into
\begin{equation}
\begin{split}
&\Gamma_{6} = \sqrt{\frac{5}{12}}\bigg| J_{z}=\pm\frac{7}{2}\bigg\rangle
+ \sqrt{\frac{7}{12}}\bigg| J_{z}=\mp\frac{1}{2}\bigg\rangle,
\\
&\Gamma_{7} = \sqrt{\frac{3}{4}}\bigg| J_{z}=\pm\frac{5}{2}\bigg\rangle
- \sqrt{\frac{1}{4}}\bigg| J_{z}=\mp\frac{3}{2}\bigg\rangle,
\\
&\Gamma_{8}^{(1)} = \sqrt{\frac{7}{12}}\bigg| J_{z}=\pm\frac{7}{2}\bigg\rangle
- \sqrt{\frac{5}{12}}\bigg| J_{z}=\mp\frac{1}{2}\bigg\rangle,
\\
&\Gamma_{8}^{(2)} = \sqrt{\frac{1}{4}}\bigg| J_{z}=\pm\frac{5}{2}\bigg\rangle
+ \sqrt{\frac{3}{4}}\bigg| J_{z}=\mp\frac{3}{2}\bigg\rangle.
    \end{split}
\label{j=7/2}
\end{equation}

\begin{figure}[t]
\includegraphics[width=8 cm]{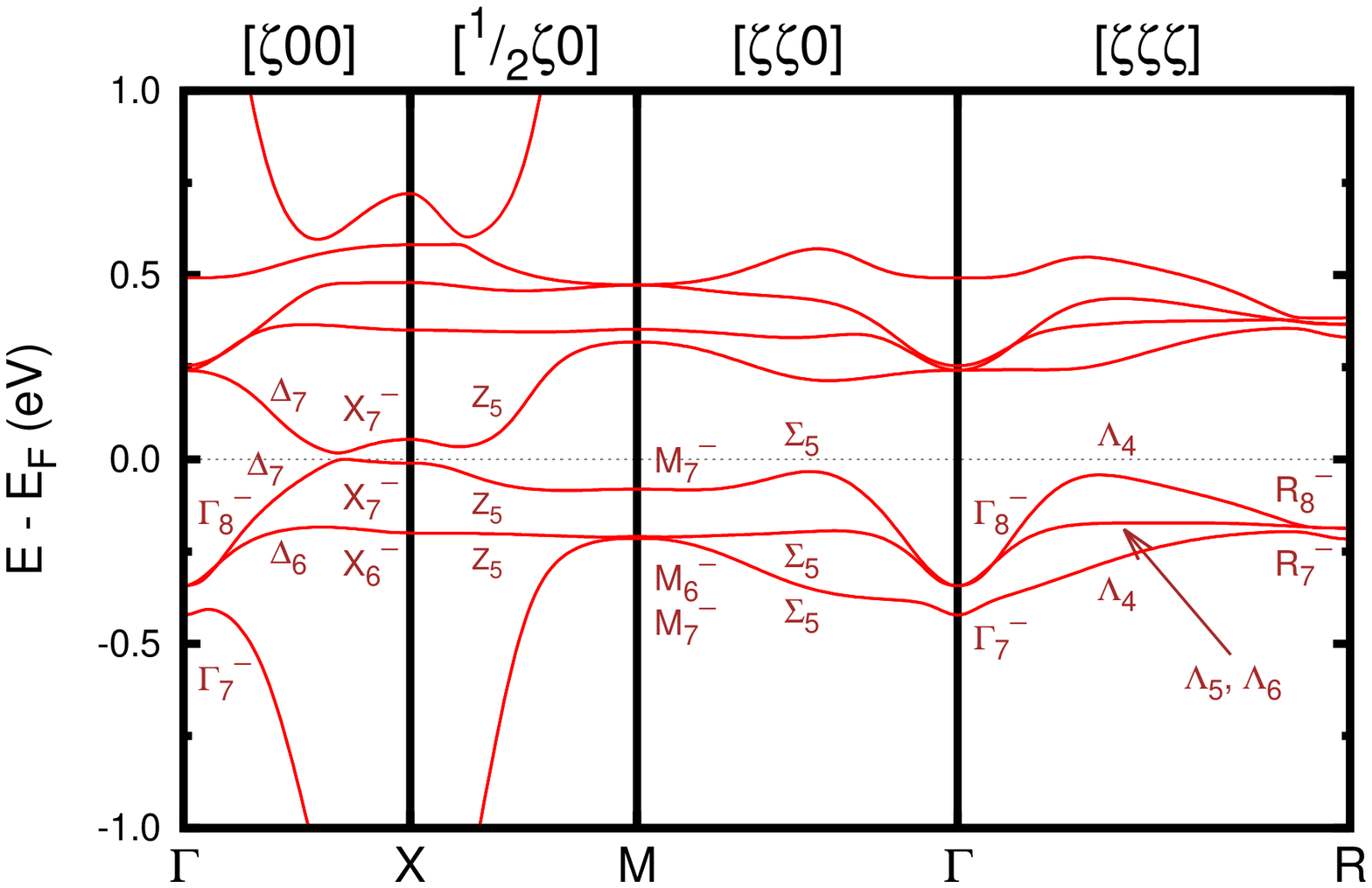}
\caption{(Color Online)
Band structure of SmB$_{6}$ in the GGA + SOC scheme.
The band symmetries at high symmetry ${\bf k}$-points are given in terms of
double group bases.
The bulk band gap of 0.018 eV and the splitting
between $J = 5/2$ and $J = 7/2$ bands of $\sim 1$ eV are obtained.
}
\label{band1}
\end{figure}

\begin{figure}[t]
\includegraphics[width=8 cm]{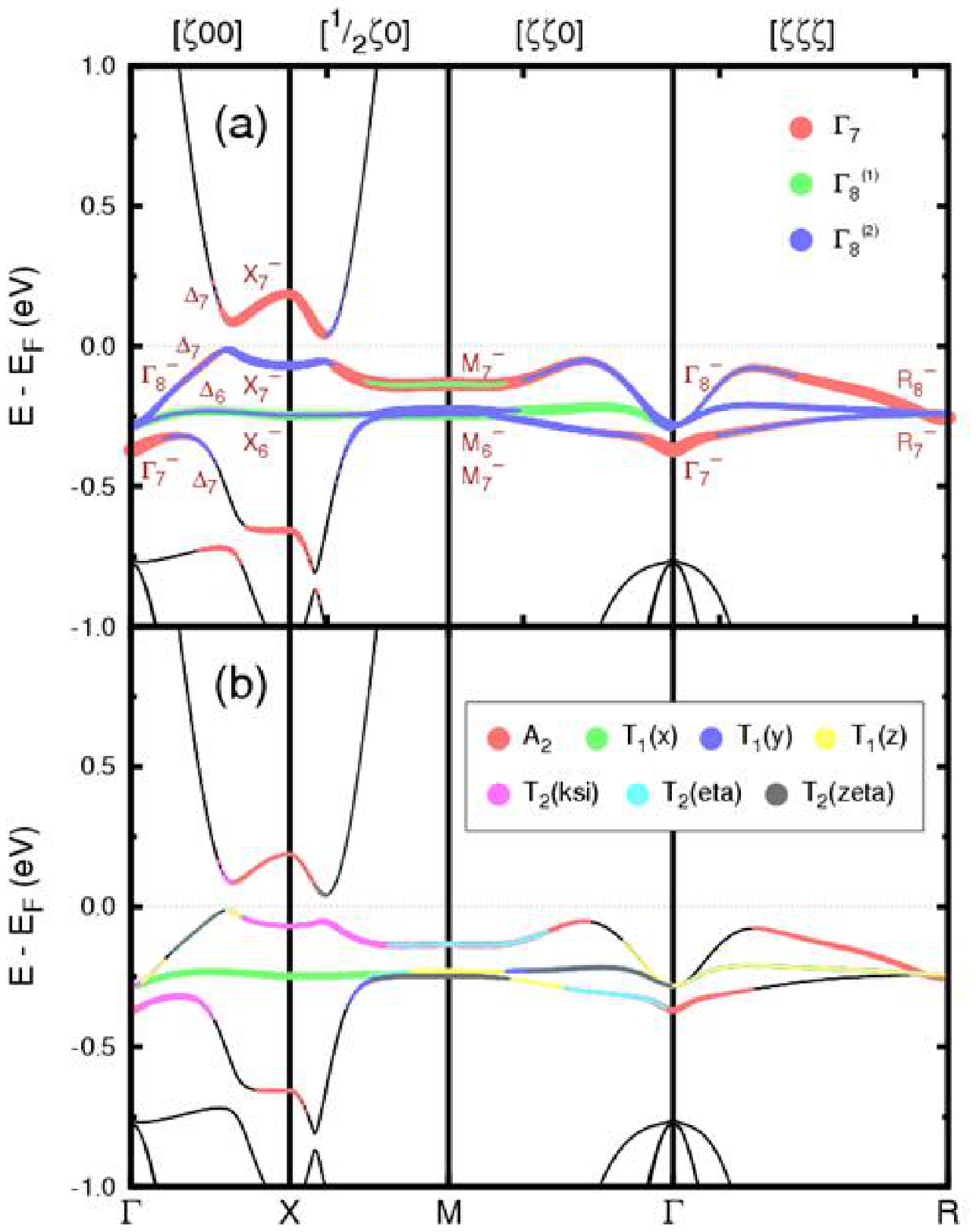}
\caption{(Color Online)
Symmetry decomposed band structures of SmB$_{6}$
with 10 times enhanced SOC of 4$f$-electron in the GGA + SOC scheme.
(a) Band symmetries of $J = 5/2$ band
in terms of relativistic double group bases.
(b) Band symmetries of $J = 5/2$ band
in terms of real cubic bases.
The bulk band gap of 0.043 eV and the splitting between $J = 5/2$ and $J = 7/2$ bands
of $\sim 5.65$ eV are obtained.
Sm $4f$ band hybridizes with Sm $5d$ band
having $e_{g}$ symmetry to produce the gap near X.
}
\label{band2}
\end{figure}

Energy levels in the $O_{h}$ cubic symmetry will be split further
when the symmetry is lowered to $D_{4h}$ tetragonal symmetry,
as shown in Fig.~\ref{cef}(b).
The crystal field splittings from $O_{h}$ to $D_{4h}$ symmetry actually occur
in the band structure along $\Gamma$-$X$ or $\Gamma$-$M$ direction,
as revealed in Figs.~\ref{band1} and \ref{band2}(a) below.

Sm $d$-electrons also feel the cubic crystal field.
Since the wave functions of $d$-electrons are spatially more spread
than those of localized $f$-electrons,
Sm $d$-electrons feel much stronger crystal field effect than the SOC effect.
Since B cages are located on every corner of the Sm-centered cube,
$e_{g}$ states that have lobes away from the anion B cages
are lower in energy than $t_{2g}$ states that have lobes
along the anion B cages, as shown in Fig.~\ref{cef}(d).

Figure~\ref{band1} shows the FLAPW band structure of SmB$_6$
in the GGA+SOC scheme.  The band gap of 0.018 eV is obtained.
It is seen that Sm $f$ bands are dominant near $E_F$, and
$J = 5/2$ and $J = 7/2$ states are below and above $E_F$, respectively.
The splitting between $J = 5/2$ and $J = 7/2$ states is about 1 eV,
which is determined by the SOC strength of 4$f$-electron in Sm atom.
We performed the basis-decomposition process for the 4$f$ bands
using Eq.~(\ref{j=5/2}).
Due to the crystal field, $J = 5/2$ band at $\Gamma$ splits into
$\Gamma_{7}^{-}$ and $\Gamma_{8}^{-}$ with the separation of about 0.1 eV.
It is seen that, at $\Gamma$, $\Gamma_{7}$ doublet is lower than $\Gamma_{8}$
quartet.  Since the lobes of $\Gamma_{7}$ are along the anion B cages,
it is tempting to conjecture that $\Gamma_{7}$ would be higher
than $\Gamma_{8}$.
The seemingly opposite situation in Fig.~\ref{band1} is expected to occur due
to the interaction between $d$ and $f$ electrons in Sm ions.
Note that the size of the wave function of $d$-electron
is much larger than that of localized $f$-electron,
and so the crystal field effect should be considered first for the $d$-electron,
which results in the lower $e_{g}$ states in energy than the $t_{2g}$ states.
Then the $f$-electron wave functions are to be arranged in a manner
to minimize the Coulomb repulsion between $d$ and $f$ electrons.
Since the lobes of $\Gamma_{8}$ and $e_{g}$ states are
along the same directions, the resulting larger Coulomb repulsion
would yield $\Gamma_{8}$ higher than $\Gamma_{7}$.

Recent band structure calculations
based on the Gutzwiller variational method~\cite{Lu13} and
the dynamical mean-field theory (DMFT)~\cite{Deng13,Jwkim13}
show that the separation between $J = 5/2$ and $J = 7/2$ band is much larger
than that of conventional DFT calculations.
This is due to the strong correlation effect of 4$f$ electrons, which
cannot be captured in the conventional DFT calculations
such as the GGA + SOC scheme.
Hence, in the conventional DFT calculations,
the splitting between $J = 5/2$ and $J = 7/2$ bands is obtained to be small
so as to bring about the unphysical overlap between them.
That is why the bands and corresponding wave functions at $\Gamma$ and $X$
tend to have wrong symmetry and shape.

To separate $J = 5/2$ and $J = 7/2$ states farther within the DFT+SOC,
we have devised a scheme to adjust the SOC strength of Sm 4$f$ electrons
artificially.
We have chosen 10 times enhanced SOC for Sm 4$f$-electron,
which gives the $J = 5/2$ and $J = 7/2$ band separation
similar to that from the DMFT.\cite{Deng13,Jwkim13}
Figure~\ref{band2} shows the band structure obtained in this scheme.
Note that $J = 5/2$ and $J = 7/2$ bands are well separated ($\sim 6$ eV),
and so the band overlap between them is highly reduced.
It is seen that Sm $d$ and B $p$ bands below $E_F$ are also slightly
shifted up.  But, apart from that, the band structure
in the vicinity of $E_F$ is almost identical to that in Fig.~\ref{band1}.
Only the band gap increases slightly from 0.018 eV to 0.043 eV.
So we can use this band structure effectively to analyze characters
and symmetries of bands near $E_F$.

\begin{figure}[t]
\includegraphics[width=7 cm]{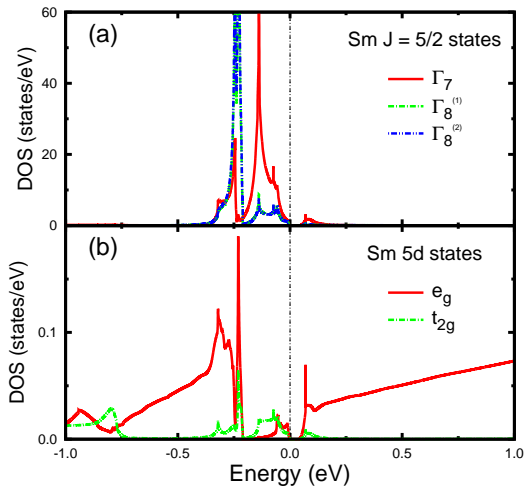}
\caption{(Color Online)
Symmetry projected partial DOSs of SmB$_{6}$ with 10 times enhanced SOC
of 4$f$-electron.
(a) Sm $J = 5/2$ DOS projected into the
relativistic double group bases.
(b) Sm $5d$ DOS projected into the real cubic bases.
}
\label{10fsoc-pdos}
\end{figure}

The crystal field order in Fig.~\ref{band2}(a) is
seen to be the same as that of the bands with normal SOC strength in Fig.~\ref{band1}.
At $\Gamma$, $J = 5/2$ splits into lower $\Gamma_{7}$ and higher $\Gamma_{8}$
with the separation of about 0.1 eV.
At X, the highest $J = 5/2$ band has the X$_{7}$ symmetry,
which has the $\Gamma_{7}$ origin (see the red colored band in Fig.~\ref{band2}(a)).
Hence, in the multiplet description, the ground state of SmB$_6$ would be $\Gamma_{7}$,
because the hole resides on the $\Gamma_{7}$ state.
This feature indicates that it is essential to take into account
the $\Gamma_{7}$-related band when modeling the band structure of SmB$_6$.
This $\Gamma_{7}$-related band is much more dispersive than the $\Gamma_{8}$-related bands.
The band width of $\Gamma_{7}$ band is as much as 0.4 eV.
Of course, the observed band width in ARPES
(10 $\sim$ 20 meV)\cite{Miyazaki12,Xu13,Neupane13,JD13}
is more than ten times smaller than the present DFT band width,
which is ascribed to the band renormalization by the
strongly correlated $4f$ electrons.

Figure~\ref{band2}(b) shows the band structure with the
real cubic basis-decomposition using Eq.~(\ref{harmonics}).
Due to the artificially enhanced SOC of 4$f$-electron,
the real cubic bases are highly mixed up with each other.
Nevertheless, it is clearly shown that the highest X$_{7}$ band has
mostly $A_2$ character,
while the flat $\Delta_{6}$ band along $\Gamma$-X at $-0.3$ eV
has mostly $T_{1}$ character.

\begin{figure}[t]
\includegraphics[width=7 cm]{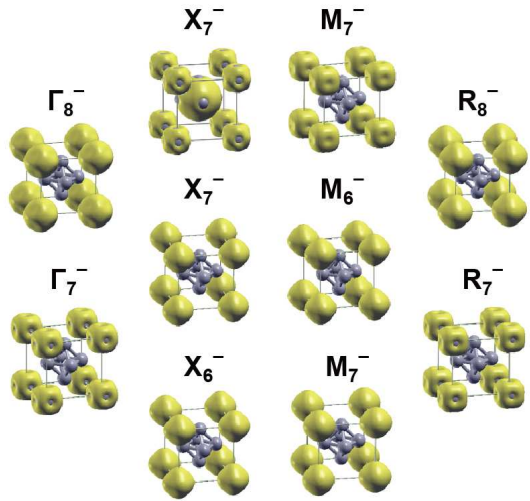}
\caption{(Color Online)
Wave functions of SmB$_6$ at high symmetry ${\bf k}$-points.
The wave function of the highest $X_{7}$ state has the contribution
from B$_{6}$ cluster, which has B 2$p$ character.
}
\label{wavef}
\end{figure}

Figure~\ref{10fsoc-pdos} shows the partial symmetry-projected
density of state (DOS) of SmB$_{6}$ with the enhanced SOC.
In Fig.~\ref{10fsoc-pdos}(a), one can notice two sharp peaks separated
by $\sim 0.1$ eV,
which correspond to mainly $\Gamma_{8}$ and $\Gamma_{7}$ of Sm $J = 5/2$ bands, respectively.
It is seen in Fig.~\ref{10fsoc-pdos}(b) that the Sm $5d$ band that overlaps
with Sm $J = 5/2$ bands corresponds to mainly $e_g$ state.
Focusing on $e_{g}$ states, $\Gamma_{8}$ states in cubic symmetry
split into $\Gamma_{6}$ and $\Gamma_{7}$,
which have mostly $d_{z^2}$ and $d_{x^2-y^2}$ component,
respectively, at X along the [001] direction in tetragonal symmetry.\cite{eg}
Hence $\Gamma_{8}$ at $\Gamma$ in Fig.~\ref{cef}(d) will be split
into $\Delta_{6}$ and $\Delta_{7}$ along X, and the latter is to be
hybridized with $\Delta_{7}$ Sm $f$-bands near $E_F$ to give rise to the energy gap.
But the latter is not to be hybridized with flat $\Delta_{6}$ Sm $f$-band,
as shown in Fig.~\ref{band2}.
Note that all the other Sm $5d$ bands are well above $E_{F}$ ($\sim$ 3 eV)
except for the $\Delta_{7}$ band,
which exhibits large dispersion of approximately 3 eV.

Figure~\ref{wavef} provides the shapes of wave functions
at high symmetry ${\bf k}$-points,
which reflect the band symmetries explicitly.
At $\Gamma$, the lower $\Gamma_{7}$ wave function has nodes
along the cubic axes, while
the upper $\Gamma_{8}$ wave function has lobes along the cubic axes.
These shapes are consistent with those of $\Gamma_{7}$,
$\Gamma_{8}^{(1)}$, and $\Gamma_{8}^{(2)}$ given in Fig.~\ref{cef}(c).
At X, the wave function of the highest $X_7$ (lowest unoccupied) band has
the $\Gamma_{7}$-like shape with nodes along the cubic axes,
while those of two lower bands ($X_7$ and $X_6$) have the $\Gamma_{8}$-like shape
with lobes along the cubic axes.
Note that the wave function of the highest $X_{7}$ has the contribution from
B$_{6}$ cluster, implying the hybridization with B 2$p$ states.
The wave functions at M and R show similar symmetries of those at X and $\Gamma$,
respectively.
This property is in agreement with the band symmetry discussed in Fig.~\ref{band2}(a).
It is highly desirable to check these band symmetries by the polarization-dependent
ARPES experiment.

\subsection{Mixed-parity semi-core band}

\begin{figure}[t]
\includegraphics[width=8 cm]{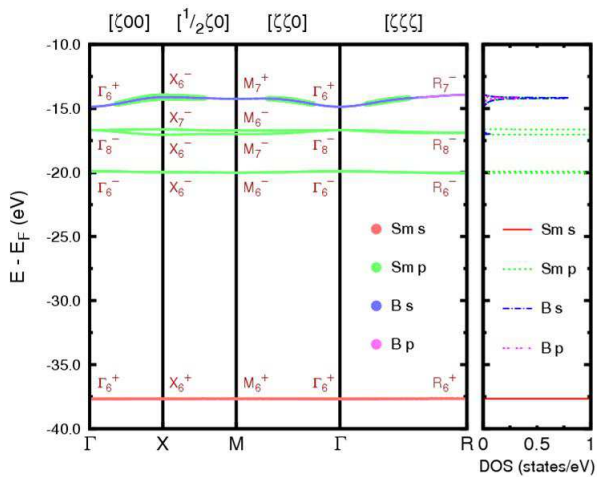}
\caption{(Color Online)
Mixed parity semi-core band of SmB$_{6}$ at $-15$ eV,
which is composed of mainly B $s$ and
partially Sm $p$ and B $p$ characters.
Below $-15$ eV, Sm $p_{3/2}$, Sm $p_{1/2}$, and Sm $s$ core bands
are shown in order from the top.
}
\label{mixedparity}
\end{figure}

We have found that the semi-core band located at $-15$ eV
has the mixed parity, as shown in Fig.~\ref{mixedparity},
which is quite unusual for the core band.
The mixed parity arises from the mixing of bands of different parities.
Namely, this mixed parity band is composed of mainly B $s$ character
and partially B $p$ and Sm $p$ characters.
Below $-15$ eV, there are Sm $p_{3/2}$, Sm $p_{1/2}$, and Sm $s$ core bands,
which are well localized (see Fig.~\ref{mixedparity}).
In contrast, the mixed parity semi-core band has some finite band width,
which is clearly seen in the DOS of Fig.~\ref{mixedparity}.
This indicates that there is some hybridization
between several atomic levels to make the band character.

\begin{table}[b]
\caption{The products of parity eigenvalues of the occupied states
(a) including core bands and (b) excluding core bands
for high symmetry ${\bf k}$-points in the bulk Brillouin zone.
Table~\ref{parity}(b) is exactly the same as ones recently reported
by Lu, \emph{et. al.}~\cite{Lu13} and Deng, \emph{et. al.}~\cite{Deng13}
}
\begin{center}
\begin{tabular}{p{2cm} p{2cm} p{2cm} p{2cm}}
\multicolumn{4}{c}{(a) Including core bands}\\
\hline \hline
$\Gamma$ & 3$X$ & 3$M$ & $R$\\
\hline
$-1$ & $-1$ & $-1$ & +1\\
\hline \hline\\
\multicolumn{4}{c}{(b) Excluding core bands}\\
\hline \hline
$\Gamma$ & 3$X$ & 3$M$ & $R$\\
\hline
+1 & $-1$ & +1 & +1\\
\hline \hline\\
\end{tabular}
\end{center}
\label{parity}
\end{table}

\begin{figure}[t]
\includegraphics[width=6 cm]{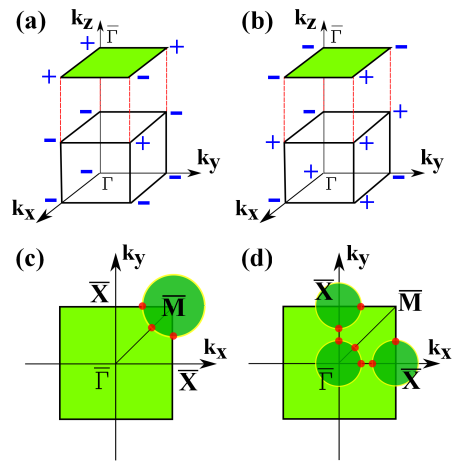}
\caption{(Color Online)
(a),(b) Parities of TRIM points in the surface Brillouin zone
are given by the product of parity eigenvalues of
underneath bulk ${\bf k}$-points.
For example, the parity (the so-called time-reversal polarization)
of $\bar{\Gamma}$ is given by the product of parities of $\Gamma$ and X.
(a) and (b) belong to TABLE.~\ref{parity}(a) and (b), respectively,
both of which yield the same topological indices of a strong topological insulator.
(c),(d) Two possible candidates of Fermi surfaces in the (001) surface Brillouin zone.
Note that they can be realized from either of two TRIM tables.
According to (a) and (b), the sign of the parity changes
along $\bar{\Gamma}\bar{M}$ and $\bar{X}\bar{M}$,
which indicates that the surface band crosses $E_F$ odd times.
In contrast, along $\bar{\Gamma}\bar{X}$,
the surface band crosses $E_F$ zero or even times.
}
\label{surface}
\end{figure}

Now, we have turned our attention into the topological symmetry in this system.
Since SmB$_{6}$ has inversion symmetry,
parity is good enough to investigate the $\mathbb{Z}_{2}$ topological number.\cite{Fu07}
If we consider all occupied states including core bands,
we obtain parity products of Table~\ref{parity}(a),
which are different from those of recent reports
by Lu \emph{et al.}~\cite{Lu13} and Deng \emph{et al.}.\cite{Deng13}
On the contrary, if we exclude core bands,
we obtained the same parity products as previous ones,
Table~\ref{parity}(b).
In fact, both parity tables give the same value of parity products, $-1$,
which corresponds to $(-1)^{\nu_{0}}$ and so
produce the nontrivial $\mathbb{Z}_{2}$ number of $\nu_{0}=1$.
Also, concerning $\nu_{0};(\nu_{1}\nu_{2}\nu_{3})$ indexation,\cite{Fu07}
these two parity tables give the same 1;(111) class,
indicating the strong topological insulator,
as shown in Fig.~\ref{surface}(a) and (b).
Therefore, despite that two parity tables have quite different
bulk parities, they give the same surface topology.
In short, the mixed-parity semi-core band gives rise to different bulk parities,
but it does not alter the surface topology in the present case.\cite{trp}
Even so, when the products of parity eigenvalues are accounted for,
it would be safer to consider all the occupied states
including the core bands.

Figure~\ref{surface}(a) and (b) show parities of TRIM points
in the surface Brillouin zone, which are obtained from
Table~\ref{parity}(a) and \ref{parity}(b), respectively.
According to Fig.~\ref{surface}(a) and (b),
the sign of the parity changes
along $\bar{\Gamma}\bar{M}$ and $\bar{X}\bar{M}$,
which implies that surface bands cross $E_F$ odd times.
Along the other directions, surface bands cross $E_F$ zero or even times.
Under these circumstances, Fermi surfaces in the surface
Brillouin zone can be realized in many different ways.
In Fig.~\ref{surface}(c) and (d),
we provide two possible candidates of Fermi surfaces
in the surface Brillouin zone.
In the former, there is one Fermi surface around $\bar{M}$,
while, in the latter, there are three Fermi surfaces,
one around $\bar{\Gamma}$ and two around $\bar{X}$.
Recent ARPES experiments~\cite{Miyazaki12,Xu13,Neupane13,JD13}
observed X and $\Gamma$-point surface states near $E_{F}$,
which corresponds to the surface topology shown in Fig.~\ref{surface}(d).
However, the surface states could be varied depending on the surface termination.
Therefore, it should be carefully checked how the surface topology changes
by varying the surface termination.

\section{Conclusion}

We have investigated band structures and
their symmetries of SmB$_{6}$ within the DFT level.
We have found the followings,
(i) at $\Gamma$, Sm $4f$ band is split into lower $\Gamma_{7}^{-}$
doublet and upper $\Gamma_{8}^{-}$ quartet,
(ii) at X, the unoccupied $4f$ band has X$_{7}^{-}$
symmetry that has $\Gamma_{7}^{-}$ origin, while
(iii) Sm $5d$ band that hybridizes with Sm $4f$ band has the
$e_{g}$ symmetry.
We have also found that the semi-core band located at about $-15$ eV
has the mixed parity due to mixed B $s$, $p$, Sm $p$ band character,
and so affects the products of parity eigenvalues.
Even though this mixed parity band does not lead to
different topological class,
this finding demonstrates that all the occupied bands including
core bands should be considered carefully to get the correct
topological symmetry.

\begin{acknowledgments}
This work was supported by the NRF (No.2009-0079947)
and the KISTI supercomputing center (No. KSC-2013-C3-010).
J.S.K. acknowledges support by the NRF (No. 2011-0022444).
J.D.D. is supported by the U.S. DOE (No. DE-AC02-05CH11231).
Helpful discussions with J. W. Allen and K. Sun are greatly appreciated.
\end{acknowledgments}


\begin{thebibliography}{99}
\bibitem{trp}
The situation here is quite similar to that in the gauge transformation.~\cite{Fu07}
Namely, the gauge transformation changes the sign of time-reversal polarization
of each TRIM point, but the product of time-reversal polarizations remains invariant.
\end{thebibliography}
\end{document}